# Effects of the difference in tube voltage of the CT scanner on dose calculation

**Dong Joo Rhee, Sung-woo Kim, Dong Hyeok Jeong**

*Medical and Radiological Physics Laboratory, Dongnam Institute of Radiological and Medical Sciences, Busan 619-953*

**Young Min Moon, Jung Ki Kim**

*Department of Radiation Oncology, Dongnam Institute of Radiological and Medical Sciences, Busan 619-953*

Computed Tomography (CT) measures the attenuation coefficient of an object and converts the value assigned to each voxel into a CT number. In radiation therapy, CT number, which is directly proportional to the linear attenuation coefficient, is required to be converted to electron density for radiation dose calculation for cancer treatment. However, if various tube voltages were applied to take the patient CT image without applying the specific CT number to electron density conversion curve, the accuracy of dose calculation would be unassured. In this study, changes in CT numbers for different materials due to change in tube voltage were demonstrated and the dose calculation errors in percentage depth dose (PDD) and a clinical case were analyzed. The maximum dose difference in PDD from TPS dose calculation and Monte Carlo simulation were 1.3 % and 1.1 % respectively when applying the same CT number to electron density conversion curve to the 80 kVp and 140 kVp images. In the clinical case, the different CT number to electron density conversion curves from 80 kVp and 140 kVp were applied to the same image and the maximum differences in mean, maximum, and minimum doses were 1.1 %, 1.2 %, 1.0 % respectively at the central region of the phantom and 0.6 %, 0.9 %, 0.8 % respectively at the peripheral region of the phantom.

Email: djrhee08@gmail.com

Fax: +82-51-720-5826

# I. INTRODUCTION

Computed Tomography (CT) measures the attenuation coefficient of an object and converts the value assigned to each voxel into a CT number [1-3], and absorbed dose to a patient in radiation therapy is calculated using the electron density of each voxel in the patient CT image. The correlation between CT number and electron density is derived from a CT scan of the materials with known electron densities and the points between the derived relations are filled up by interpolations. This relation is applied during the dose calculation based on CT image and thus, the accurate measurement of CT number and applying the value to the treatment planning system (TPS) is fundamental to the radiation therapy. Attenuation coefficient varies with a chosen X-ray tube voltage (kVp) from the CT scanner since attenuation coefficient has an energy dependent property and many literatures [4-6] have discovered the significant CT number changes in different kVp for most materials and therefore, application of the correct CT number to electron density conversion curve is essential. However, in practice, many institutions use various tube voltages for CT images depending upon the section of the body or the size of the patients but often, a single CT number to electron density conversion curve is utilized for the radiation therapy plan. Although the uncertainties in dose calculation caused by the various CT tube voltages have been quantified in some literatures by applying different conversion curves [7], the dose was calculated only with the TPS with box phantoms in a fixed condition. In this study, firstly, the CT number to electron density conversion curves for each tube voltage would be derived to see if they have any statistically significant variations. Then, the errors in percentage depth dose (PDD) using TPS and Monte Carlo methods were calculated and evaluated. Finally, a single 3D clinical case using the head part of the anthropomorphic phantom would be analyzed to quantify the inaccuracy in dose calculation by the wrong usage of the conversion curve.

## II. EXPERIMENTS

An attenuation coefficient µ of a material is defined as

$$\mu = \sigma N \tag{1}$$

where $\sigma$ is the total cross-section of the material and $N$ is the number of atoms per unit volume. The total cross-section of a material is the sum of the cross-section of the all possible interactions including photo-electric interaction and Compton scattering interaction for keV range of photons. The cross-section values in this range is energy dependent and therefore, the CT number which is defined to be

$$CT\ number = 1000 \times \frac{\mu - \mu_{water}}{\mu_{water}} \tag{2}$$

becomes energy dependent, where $\mu_{water}$ is the attenuation coefficient of water.

GE (General Electric Healthcare, Waukesha, WI) Lightspeed CT scanner was used to acquire the data in this study. The beam current was 330 mA and the slice thickness and the pixel size were 2.5 mm and 0.62 mm respectively. The field of view (FOV) of the CT image was 512 x 512 and standard kernel was used to reconstruct the image. The tube voltage of 80 kVp, 100 kVp, 120 kVp, and 140 kVp were tested to evaluate the impact on the CT number by the various tube voltages.

The sensitometry of the image quality verification phantom (Catphan 504, Phantom Laboratories, Salem, NY) was used to gain the CT number to electron density conversion curves for each tube voltage as shown in fig. 1. In the sensitometry, 7 different materials were placed and the relative electron density of each material (acrylic 1.147, air 0.001, Delrin 1.363, LDPE 0.945, PMP 0.853, polystyrene 0.998, and Teflon 1.868 when the electron density of water was assumed to be 1.000) given in the data sheet offered by the manufacturer was used to produce the CT number to electron density conversion curve. The relations between the CT number and the electron density of the materials were also derived for dose calculation. The CT number was determined to be the average of square of 49 pixels, locating at the center of each material cylinder as shown in fig. 1. The sensitometry CT image was taken 5 times to derive the standard error of the CT numbers in the materials.

Percentage depth dose (PDD) calculation was performed on the cubic water phantom, 30 cm each side. The phantom was computationally generated and had the relative electron density of 1.0 uniformly. The phantom images were assumed to be taken with the tube voltage of 80 kVp and 140 kVp and the CT number to electron density curve from 80 kVp was assumed to be applied for both images. Therefore, the electron density of the phantom was correctly assigned to be 1.00 for the 80 kVp image, whereas the phantom density was assumed to be 1.04 for the 140 kVp image. This electron density overestimation was due to the 4 % discrepancy in CT number between the 80 kVp and 140 kVp conversion curves at electron density of 1.0 according to fig. 2.

6 MV anterior-posterior (AP) direction circular shape photon beam with a diameter of 6 cm was irradiated to the surface of the phantom when the field size was defined at source to axis distance (SAD) of 100 cm. The center of the cubic phantom was 100 cm away from the beam source and located at the beam central axis. Dose calculation was performed with the two different methods: equivalent tissue-air ratio method (ETAR) algorithms from CorePLAN (Seoul C&J, Seoul, South Korea) treatment planning system, and Monte Carlo simulation using Geant4 version 9.4 patch 02 particle simulation toolkit [8,9]. Previously modeled Varian's Clinac 2100C/D 6 MV beam energy spectrum [10] was adapted to the Monte Carlo simulation.

To quantify the clinical impact from the variations in tube voltages, a simple 3D radiotherapy plan was performed on the head part of the anthropomorphic phantom (RANDO Phantom, Phantom Laboratories, Salem, NY) with ETAR algorithm. The CT image of the phantom was taken with 80 kVp. The two different CT number to electron density conversion curves were applied (80 kVp and 140 kVp in fig. 2) to the image taken with 80 kVp and each plan was called 80 kVp plan and 140 kVp plan. Therefore, the 80 kVp plan became the reference plan. 4 beams (anterior-posterior, posterior-anterior, right lateral, left lateral) that have the field sizes covering the cylindrical volume at the central region of the phantom head with 3 mm margin. Furthermore, four cylindrical volumes were located at the peripheral regions of the head where the beams enter. All volumes were called volume of interest (VOI) and their positions were shown in fig. 3. The 80 kVp reference plan was normalized to the central VOI mean dose of 200 cGy and the beams with same MUs were identically delivered to the 140 kVp case. Mean, maximum, and minimum doses to the each VOI were calculated for the 80 kVp and 140 kVp plans.

## III. RESULTS AND DISCUSSION

The error bar in fig. 4 represented the standard error of 2-sigma and CT number variations of the different materials for various kVps were statistically significant as shown in fig. 4 except for air, where error bars overlapped each other apart from the 80 kVp case. This is because the electron density of air was extremely low and consequently, the air CT image became very sensitive to the noise so the CT number variation caused by imaging noise was dominant over the tube voltage effects. CT number linearly increased with increment in kVp except for Teflon, and the results was consistent with the other literature [4]. The maximum differences were shown between the values from 80 kVp and 140 kVp as expected. The differences were from 18.3 HU (Delrin) to 52.9 HU (Teflon) and the standard errors were minuscule being compared with the differences in CT numbers between each kVps.

Percentage depth dose (PDD) curves with the phantom image from the two different tube voltages (80 kVp and 140 kVp) with applying the same CT number to electron density were derived with the CorePlan TPS and Geant4 simulation as demonstrated in fig. 5 (upper left and lower left). From the subtractions between two curves in fig. 5 (upper right and lower right), 140 kVp PDD was higher at the build-up region and then 80 kVp PDD became higher after the maximum depth dose. The results seemed to be reasonable as the CT number of water for 140 kVp was higher than that of 80 kVp according to fig. 2 and this occurred the earlier reaching to the dose maximum depth and the faster photon attenuation afterwards. The maximum difference in dose was 1.3 % for the TPS calculation with ETAR algorithm and 1.1 % for Geant4 simulation and as the PDD differences between the two algorithms was known to be the order of 1 % - 4 % in a clinical case [11], the difference in this study was within the acceptable range.

The mean, maximum, and minimum doses in the central VOI and the 4 peripheral VOIs with two different CT number to electron density curves application were demonstrated in table 1. The dose differences in ~~tumor~~ the central VOI were 1.1 %, 1.2 %, and 1.0 % for mean, maximum, and minimum doses respectively and that of the peripheral VOI were 0.6 %, 0.9 %, and 0.8 % respectively at most. The differences were higher for the central VOI because 4 beams were overlapped in the central VOI so the differences in dose by each beam were accumulated. On the other hand, only two beams were overlapped in the peripheral VOI so the overall accumulation effect in the peripheral region was comparatively small.

## IV. CONCLUSION

Overall, the difference in CT number by different kVps was statistically significant for most materials except for air where the density is very low and therefore, noise affects the image more significantly than the cross-section differences from various photon energies. The maximum dose difference caused by applying a wrong CT number to electron density conversion curve was expected to be more than 1 % for both PDD measurement and a clinical case and the difference was likely to be significant at the region where many beams were overlapped. Although many researches have been conducted to reduce the delivered dose error, reducing 1 % error in prescription dose by simple adjustment must be preceded before applying a high-end complex technology. Therefore, therapists and medical physicists who perform the CT operation and radiation therapy planning should acknowledge each other if the different tube voltage would be used in the process of taking a CT image to reduce the possible error.

ACKNOWLEDGEMENT

This work was supported by the National Research Foundation of Korea(DIRAMS) grant funded by the Korea government(MSIP) (50495-2015)

Table 1. Mean, maximum, and minimum doses of each VOI and the tumor for 80 kVp and 140 kVp cases and their percentage differences.

| 80 kVp (Dose in cGy) | | | 140 kVp (Dose in cGy) | | | Percentage difference |
|---|---|---|---|---|---|---|
| Central VOI (Red) | Mean | 200.0 | Central VOI (Red) | Mean | 202.1 | -1.1 |
| | Max | 220.9 | | Max | 223.5 | -1.2 |
| | Min | 162.9 | | Min | 164.5 | -1.0 |
| Right VOI (Blue) | Mean | 113.5 | Right VOI (Blue) | Mean | 114.2 | -0.6 |
| | Max | 128.0 | | Max | 128.9 | -0.7 |
| | Min | 86.8 | | Min | 87.4 | -0.7 |
| Left VOI (Purple) | Mean | 111.2 | Left VOI (Purple) | Mean | 111.8 | -0.5 |
| | Max | 125.9 | | Max | 126.8 | -0.7 |
| | Min | 78.6 | | Min | 79.0 | -0.5 |
| Anterior VOI (Green) | Mean | 106.3 | Anterior VOI (Green) | Mean | 106.9 | -0.6 |
| | Max | 121.5 | | Max | 122.3 | -0.7 |
| | Min | 74.1 | | Min | 74.7 | -0.8 |
| Posterior VOI (Yellow) | Mean | 102.7 | Posterior VOI (Yellow) | Mean | 103.3 | -0.6 |
| | Max | 117.5 | | Max | 118.5 | -0.9 |
| | Min | 73.6 | | Min | 73.6 | 0.0 |

Figure Captions

Fig. 1. CATPhan sensitometry CT image with various materials, PMP, Air, Teflon, Delrin, Acrylic, Air, Polystyrene, and LDPE in order of number increasing.

Fig. 2. CT number to electron density conversion curves for 80, 100, 120, and 140 kVp cases.

Fig. 3. A slice of the CT image of the anthropomorphic phantom with contours. The tumor (red) is located at the center of the image and 4 VOI (blue, yellow, purple, and green circles) are located at the entrance of the beam.

Fig. 4. CT number variations according to change in kVp for 7 different materials.

Fig. 5. PDD curve for 80 and 140 kVp cases from Monte Carlo simulation (upper left) and TPS dose calculation (lower left) and their differences (upper right and lower right)

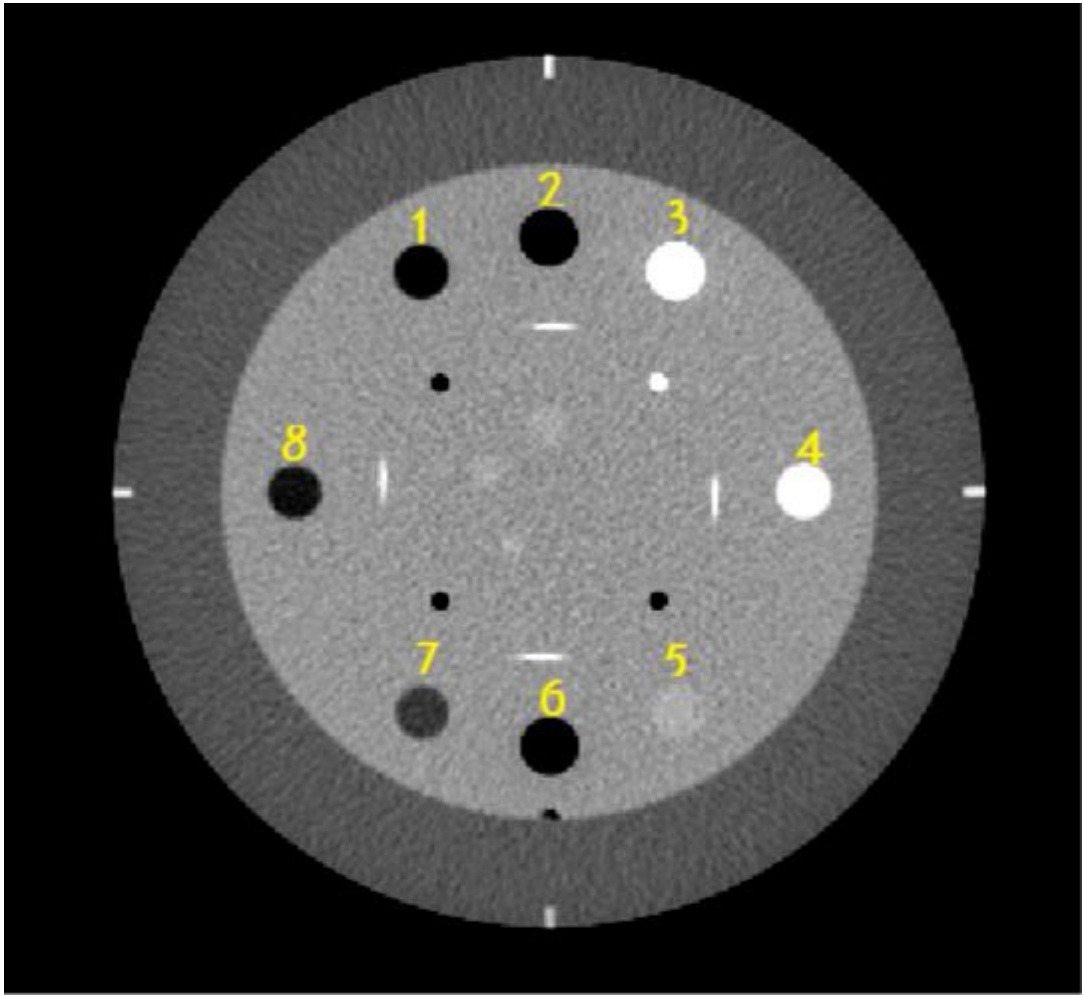

Fig. 1.

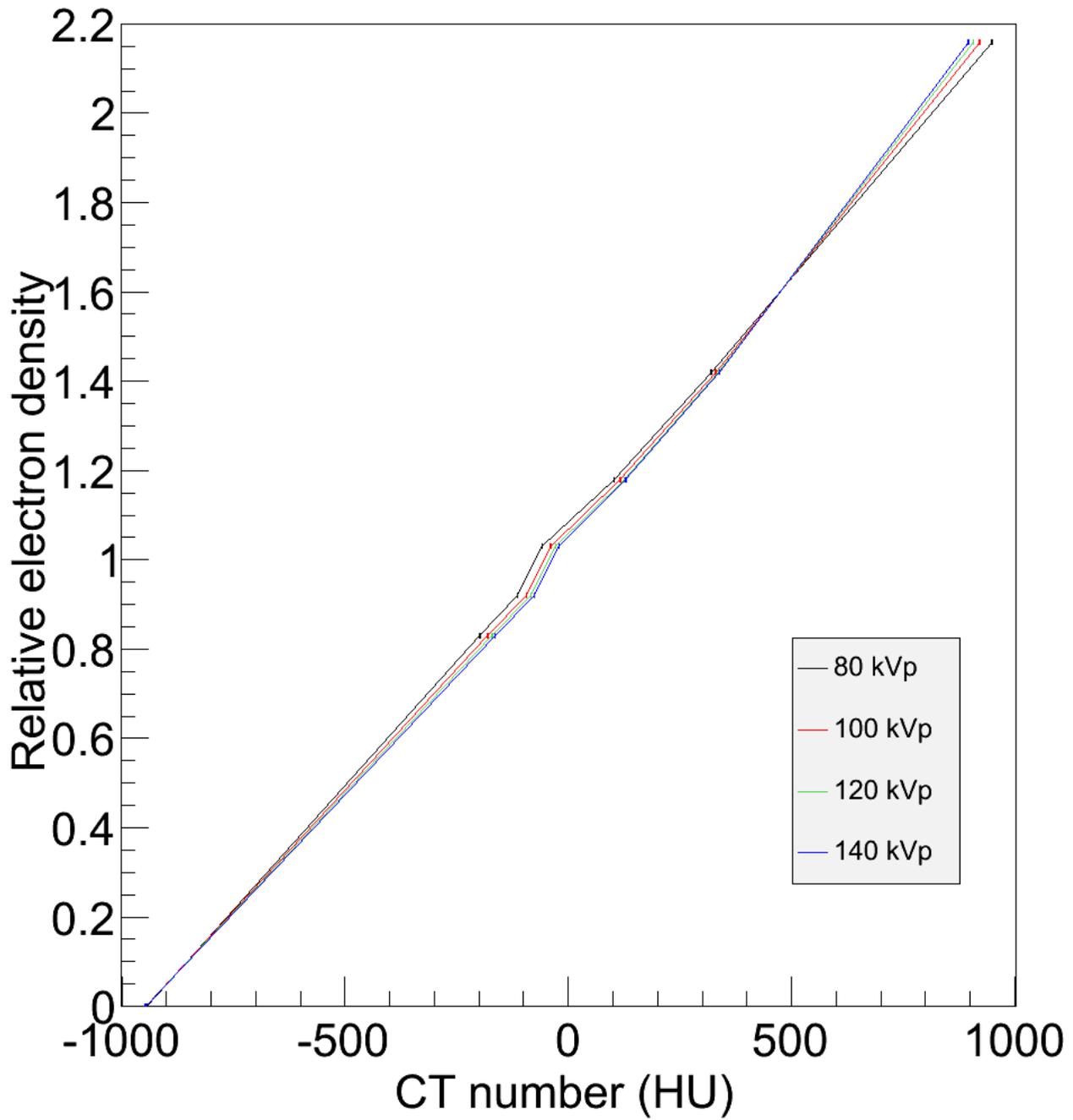

Fig. 2.

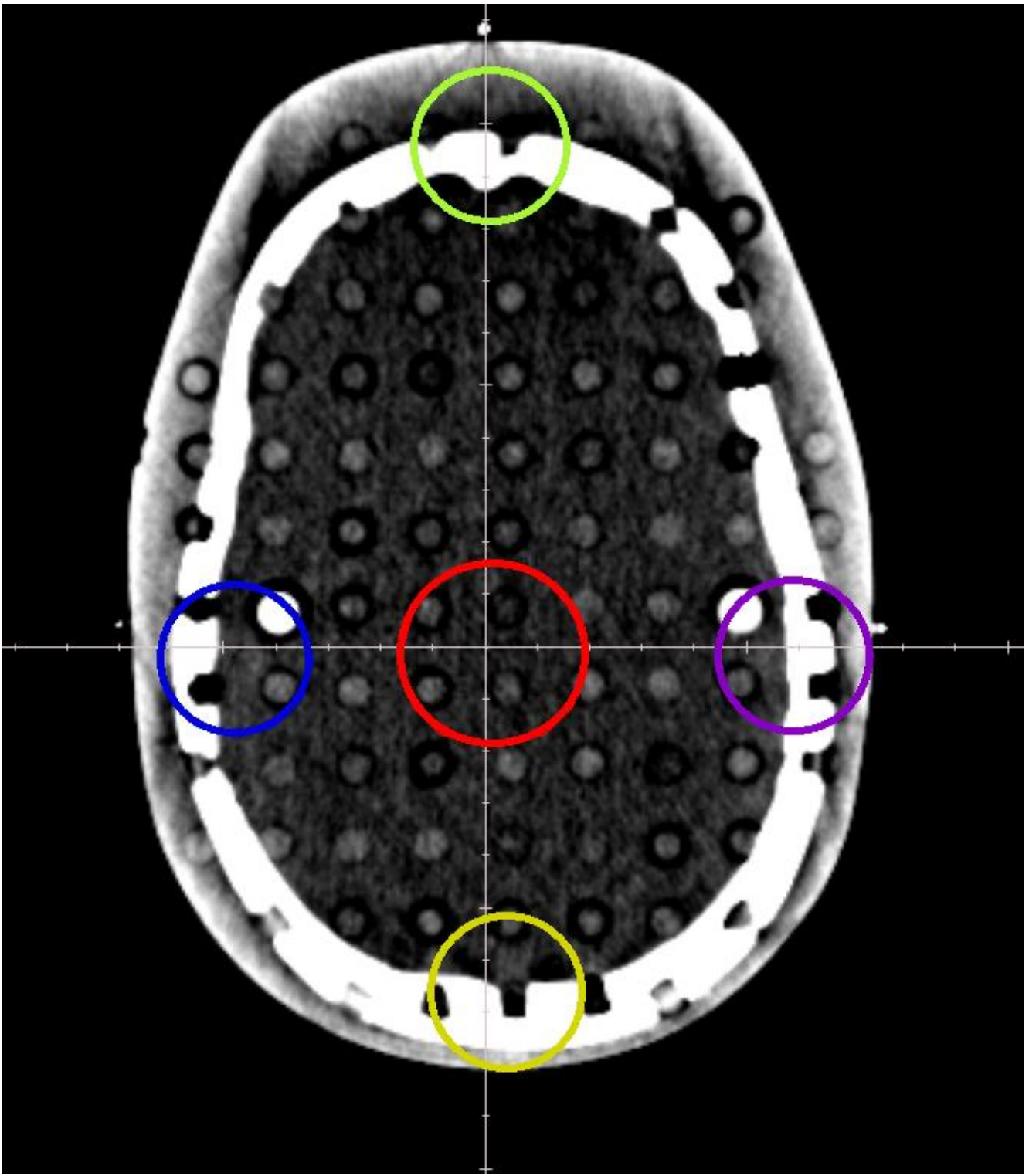

Fig. 3.

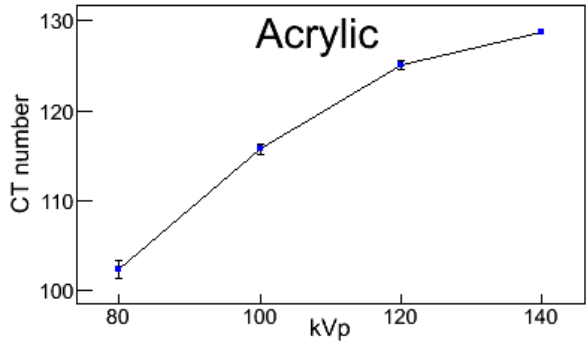
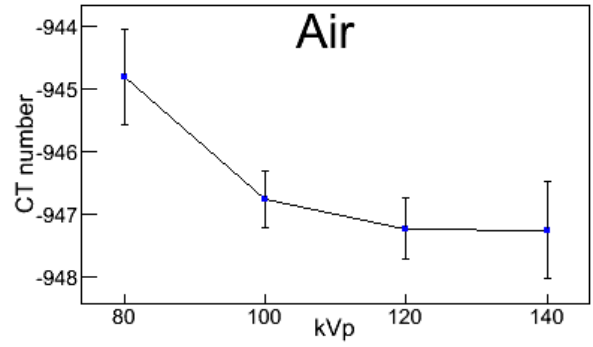
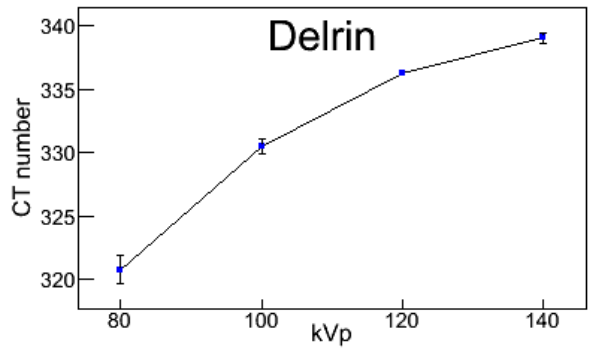
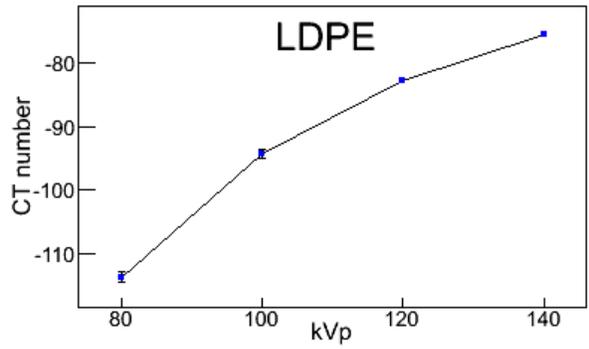
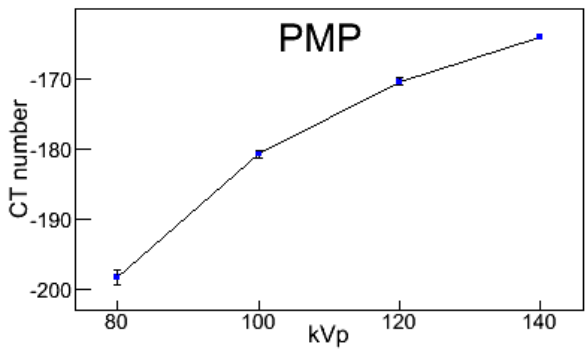
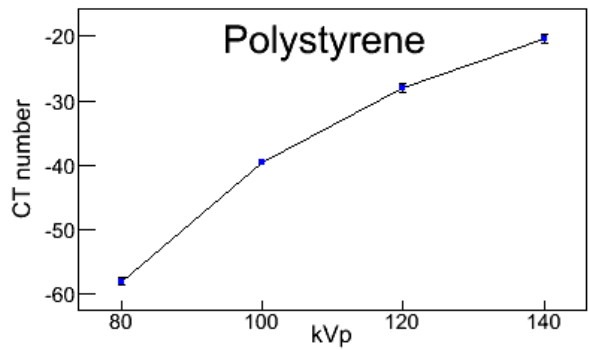
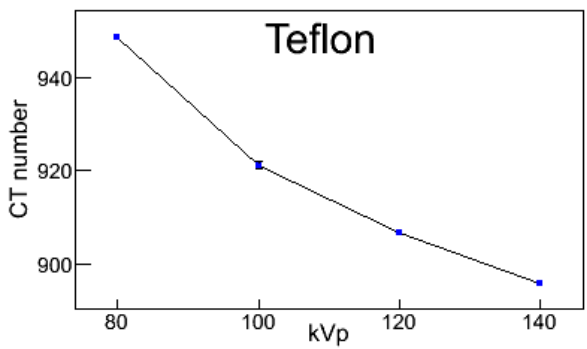

Fig. 4.

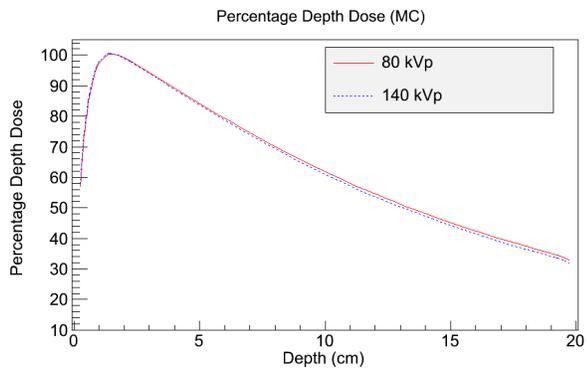
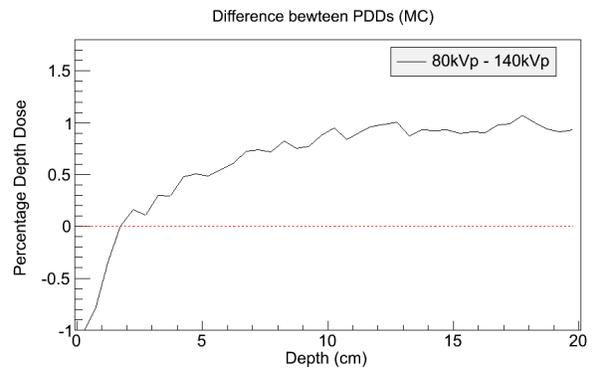
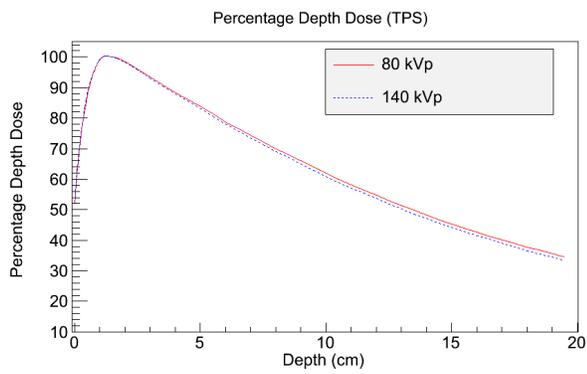
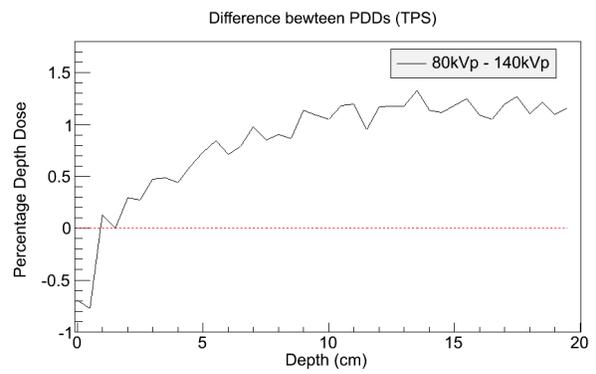

Fig. 5.